\documentclass[runningheads]{llncs}

\usepackage[T1]{fontenc}    
\usepackage{graphicx}      
\usepackage[colorlinks=true, allcolors=blue]{hyperref}  
\usepackage{color}

\usepackage{float}         
\usepackage{url}           
\usepackage{array}  
\usepackage{comment}
\usepackage{caption}

\begin{document}

\title{The Composition of Digital Twins for Systems-of-Systems: a Systematic Literature Review}
\titlerunning{Composition of Digital Twins for Systems-of-Systems}

\author{Mennatullah T. Khedr\inst{1}\orcidID{0009-0007-0096-3375} \and
John S. Fitzgerald\inst{1}\orcidID{0000-0001-7041-1807}}

\authorrunning{M T Khedr and J S Fitzgerald}

\institute{School of Computing, Newcastle University, UK \\
\email{m.t.y.a.khedr2@newcastle.ac.uk, john.fitzgerald@newcastle.ac.uk}
}

\maketitle

\begin{abstract}
Digital Twins (DTs) are increasingly used to model complex systems, especially in Cyber-Physical Systems (CPS) and System-of-Systems (SoS), where effective integration is key. This systematic literature review investigates DT composition and verification and validation (V\&V) methodologies. Analyzing 21 studies from 2022-2024, we examined composition mechanisms, SoS characteristics, and V\&V formality, scope, and challenges. While composition is discussed, formalization is limited. V\&V approaches vary, with semi-formal methods and simulations dominating; formal verification is underutilized. Key technical challenges include model uncertainty and integration complexity. Methodological challenges highlight the lack of standardized DT-specific V\&V frameworks. There is a need to move beyond model validation to address integration and cyber-physical consistency. This review contributes a structured classification of V\&V approaches and emphasizes the need for standardized, scalable V\&V and rigorous composition methodologies for complex DT implementations.
\end{abstract}

\section{Introduction}
\label{sec:intro}

Digital Twins~(DTs) are attracting widespread interest as a response to the challenges and opportunities of digitalisation in many domains. As reliance on Cyber-Physical Systems (CPSs) and Systems of Systems (SoSs) grows, so does the need for DT engineering methods and tools that allow the validation and verification (V\&V) of key qualities of DTs and DT-enabled systems. There is a growing body of work on DT architectures, design frameworks and reference models, and evaluation methods to ensure dependability and efficiency. However, there is as yet little work on V\&V for DTs in the context of cyber-physical SoSs.  

We argue that a Systematic Literature Review (SLR) is needed to guide research and innovation in this area. We here describe such a review, addressing trends in integration patterns for DT-enabled systems, system-level properties and desired qualities, techniques being addressed in V\&V, and the challenges identified so far. To promote transparency and rigour, we follow the Preferred Reporting Items for Systematic Reviews and Meta-Analyses~(PRISMA) guidelines~\cite{Pagen71PRISMA}-- a transparent framework for conducting and reporting SLRs that is increasingly applied in engineering and computing. The background, problem statement, objectives and Research Questions~(RQs) are introduced in Section~\ref{sec:context}. The review methodology is described~(Section~\ref{sec:method}) with results to date and discussion of answers to the RQs~(Section~\ref{sec:results&disscussion}). Threats to validity are discussed in Section~\ref{sec:threats} and conclusions are presented in Section~\ref{sec:conclusion}.

\section{Context}
\label{sec:context}

\subsection{Background}
\label{sec:background}
Our work is at the intersection of CPS, SoS and DT engineering, and so we briefly address each of these areas. 

CPSs are composed of computational and physical processes interacting in a defined framework~\cite{incose2023handbook}. They arise in domains such as smart grids, industrial control and autonomous vehicles where high degrees of automation, distributed control, and networked communication are required \cite{Kalyvas2021,Samak2023}. The dependable development, operation and maintenance of CPSs faces challenges posed by their heterogeneity, the complexity of their interactions, and the need to operate in dynamic, uncertain environments. Further, some applications (e.g., in logistics or the built environment) demand collaboration between CPSs, creating cyber-physical SoSs, further accentuating these challenges. 

DTs offer a way to address complexities in CPS engineering. A DT contains a virtual representation of a CPS which is often simply termed the \emph{Physical Twin}~(PT)~\cite{Fitzgerald2019}. At the core of the DT is an abstract model (or collection of models) describing aspects of the PT that are relevant to the DT's purpose and context~\cite{lv2022DTsBook}. Bi-directional data and control flows keep the DT and PT consistent~\cite{EstGomm2021}. The DT adds value by offering analysis, simulation, and optimization services, enabling performance enhancement, fault prediction, and decision-making. As interest in DT technology grows rapidly, the development of a discipline of DT engineering is naturally rather slower~\cite{Fitzgerald2024}, and it is noteworthy that there is currently a lack of standardized procedures or objectives for V\&V of DTs~\cite{Bitencourt2025V&VSLR}.

SoSs are composed of independently owned or managed constituent systems~\cite{incose2023handbook}. SoS engineering must therefore handle the autonomy, heterogeneity, and emergent behaviour of these constituents as they evolve. DTs are an attractive way to achieve this~\cite{Olsson2023}. Existing research offers frameworks for SoS modelling~\cite{Mandel2022}, formal languages and methodologies for SoS engineering~\cite{Lana2019,Nguyen2019}, and verification~\cite{Seo2016}. However, a lack of systematic methods for composing DTs hampers the full realisation of DT-enabled SoSs. Research suggests that existing DT frameworks struggle with reuse and modularity, making it difficult to integrate DTs from constituent systems into larger SoS structures effectively~\cite{JudithMichael2022}, although recent work addresses automated DT composition, aiming to minimize manual intervention and enhance adaptability~\cite{Gill2024Pipeline}.

Researchers have studied the challenges and knowledge gaps for efficient and sustainable DTs \cite{JudithMichael2022,Borth2019}, reference models for DT specifications and requirements \cite{Moyne2020,Diakité2023,BOYES2022103763}, DT design frameworks (conceptual architectures) \cite{Piras2024}, and DT models. Although there have been several publications on DT creation, they tend either to be at a general level without insight into pragmatics, or are domain-specific. 

Few recent publications have investigated the state of the art of DTs in the context of SoSs~\cite{Olsson2023,Diakité2023}. Some have addressed challenges that emerge from the integration of DTs into SoSs. For example, Cavalcante and Batista \cite{EvertonCavalante2024} offer three architectures that could result from such integration, each defining the complexity and limitations of the composed DT. These include: a global DT influencing an entire SoS; virtual replicas of each constituent system so that a DT may influence individual constituents; unique DTs for each constituent.  There is also some discussion of horizontal integration of DT parts, where the system models within the SoS have multiple views of the PT~\cite{JudithMichael2022}.

\subsection{Problem Statement, Objective and Research Questions}
\label{sec:objective}

An informal consideration of the background outlined in Section~\ref{sec:background} suggests a gap regarding formal V\&V frameworks for DTs operating in SoS environments. V\&V are crucial aspects of DT development~\cite{Perisic2024Helix}, so the gap is noteworthy given the increasing implementation of DTs across interconnected, complex systems where traditional validation approaches may prove insufficient. There is limited work on consolidating approaches to the development and deployment of DTs across domains \cite{Fett2023} and few publications have addressed evaluation methods to ensure the dependability and efficiency of the proposed frameworks and models~\cite{yao2023systematic}. Although there has been some systematic review of V\&V for DTs in manufacturing~\cite{Bitencourt2025V&VSLR}, we still lack a common framework across domains. 

The objective of this SLR is to address this gap by investigating V\&V practices for DTs in the broad context of cyber-physical SoSs. To address this, we proposed the following research questions to guide our review:
\begin{itemize}
    \item \textbf{RQ1:} What are the current approaches and integration patterns used in the composition of DTs within SoS contexts?
    \item \textbf{RQ2:} What system-level properties and quality attributes of DTs are addressed in studies focusing on their V\&V?
    \item \textbf{RQ3:} What are the existing approaches for verifying and validating DTs within SoS contexts?
    \item \textbf{RQ4:} What challenges are identified in the verification and validation of DTs within SoS environments?
\end{itemize}
To address RQ1, the main application domains discussing DT composition and the trending composition approaches were identified, noting SoS characteristics influencing the integration pattern. 
For RQ2, DT properties and qualities were first accumulated, and publications were then categorized accordingly. 
For RQ3, papers that addressed verification and/or validation of DTs in SoS context were considered, not those using DTs to verify/validate the twinned systems. 
Finally, RQ4 seeks to identify the challenges that face  V\&V methods for composed DTs.

\section{Methodology}
\label{sec:method}

A structured review protocol was established following Kitchenham's guidelines~\cite{Kitchenham2004SLR} to help ensure rigour, transparency and reproducibility. The four RQs above lead to a search strategy, inclusion and exclusion criteria, and quality assessment methods to extract the most recent relevant knowledge. Results were reported following PRISMA guidelines.

\subsection{Data Sources and Search Strategy}
To ensure comprehensive coverage, we used multiple digital libraries and indexing databases. The ACM Digital Library, IEEE Xplore, Scopus, Web of Science and Engineering Village were selected due to their extensive coverage of research in CPSs, SoSs and DTs. In addition, manual search and forward citation searching were undertaken to support the findings and reduce the risk of missing relevant recent publications. These additional records were chosen based on initial title and abstract screening.

We used the Advanced Search option in each database and library with the goal of reducing false positives. Search terms were applied using Boolean operators to ensure precision. Database-specific formatting of queries was considered to yield the best results because of the support for such features as exact phrases and wildcards. The query was structured as follows:
\begin{verbatim}
    Title: ("digital twin*")
    AND Abstract: (("cyber physical" OR "system* of systems" OR 
                    "complex system*") AND (verif* OR validat*))
    AND Keywords: ("digital twin*" OR "cyber physical" OR 
                   "system* of systems" OR "complex system*" OR 
                   verification OR validation OR "formal")
\end{verbatim}

\subsection{Inclusion and Exclusion Criteria}
Only publications available to the authors through open or institutional access could be included in the SLR. To ensure relevance and quality of selected publications, inclusion and exclusion criteria were defined. 

\textbf{Inclusion Criteria:}
Only peer-reviewed journal articles, conference papers, and book chapters were included. Search was limited to publications in English. The timespan was initially left open but was later restricted to 2021 onwards due to the lack of earlier relevant papers. Studies explicitly discussing verification and/or validation of DTs of CPSs, SoSs, or complex systems were included, as were publications on DT composition. 

\textbf{Exclusion Criteria:}
Studies that did not discuss DTs as the main topic were excluded. Publications developing a new DT architecture, framework, or application without discussing verification or validation were marked irrelevant. Duplicates across multiple databases were removed.

\subsection{Selection Process}
\label{sec:selection}
The search, conducted in February 2025, yielded 390 publication records. After deduplication, 193 entered screening. All publications were tagged during screening, making it easier to extract answers for the RQs. References and notes were organized and managed using the EndNote 21 Desktop application.

In initial screening, titles and abstracts were reviewed based on the inclusion and exclusion criteria. Publications were classed as \emph{Relevant}, \emph{Maybe Relevant}, or \emph{Irrelevant}. The \emph{Relevant} category included papers with abstracts that clearly discussed composition of DTs for SoSs or discussed V\&V of DTs. Publications considered \emph{Maybe Relevant} included those addressing supporting knowledge, e.g., formal methods, DT requirements or qualities analysis, or SoS foundations. Papers that only proposed a DT for a new application, a new DT service or cybersecurity of DTs were excluded and classed \emph{Irrelevant}. Fewer than half the initial references passed the initial screening of the title and abstract.

The \emph{Relevant} and \emph{Maybe Relevant} publications underwent full-text reading and assessment for relevance and methodological rigour. A further 36 papers were selected from manual search and forward citation searching of these potentially relevant publications. They were chosen based on title and abstract and classified into one of the three groups. This led to a total of 62 \emph{Relevant}, 36 \emph{Maybe Relevant}, and 113 \emph{Irrelevant} publications. Studies passing all the inclusion and exclusion checks were passed on to data extraction and synthesis.

\begin{figure}[htbp]
    \centering
    \begin{minipage}{0.35\textwidth}
    \centering
        \begin{tabular}{@{}
            >{\raggedright\arraybackslash}p{3cm}
            >{\centering\arraybackslash}p{1.3cm}
        @{}} \hline
            Data Source& \#Papers\\
            \hline
            Scopus& 164\\
            Web of Science& 120\\
            IEEE Xplore& 43\\
            ACM Digital Library& 35\\
            Engineering Village& 28\\
            \hline
        \end{tabular}
        \captionof{table}{Number of publications per data source}
        \label{tab:publicationsPerSource}
   \end{minipage}
    \hfill
    \begin{minipage}{0.6\textwidth}
        \centering
        \includegraphics[width=1\textwidth]{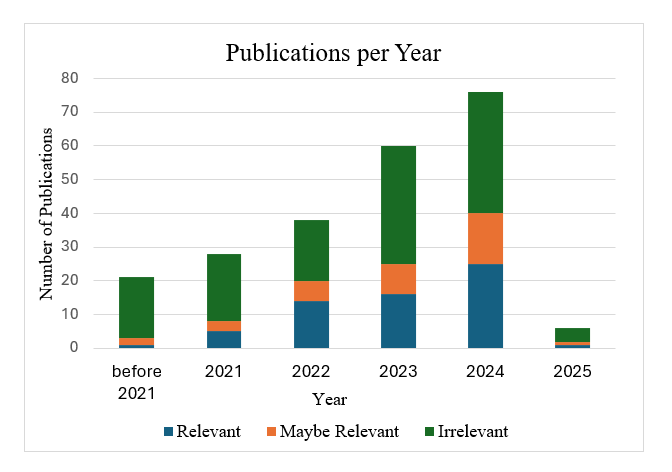}
        \caption{Number of publications per year}
        \label{fig:publicationsPerYear}
    \end{minipage}
\end{figure}

\begin{figure}[htbp] 
    \centering
    \includegraphics[width=\textwidth]{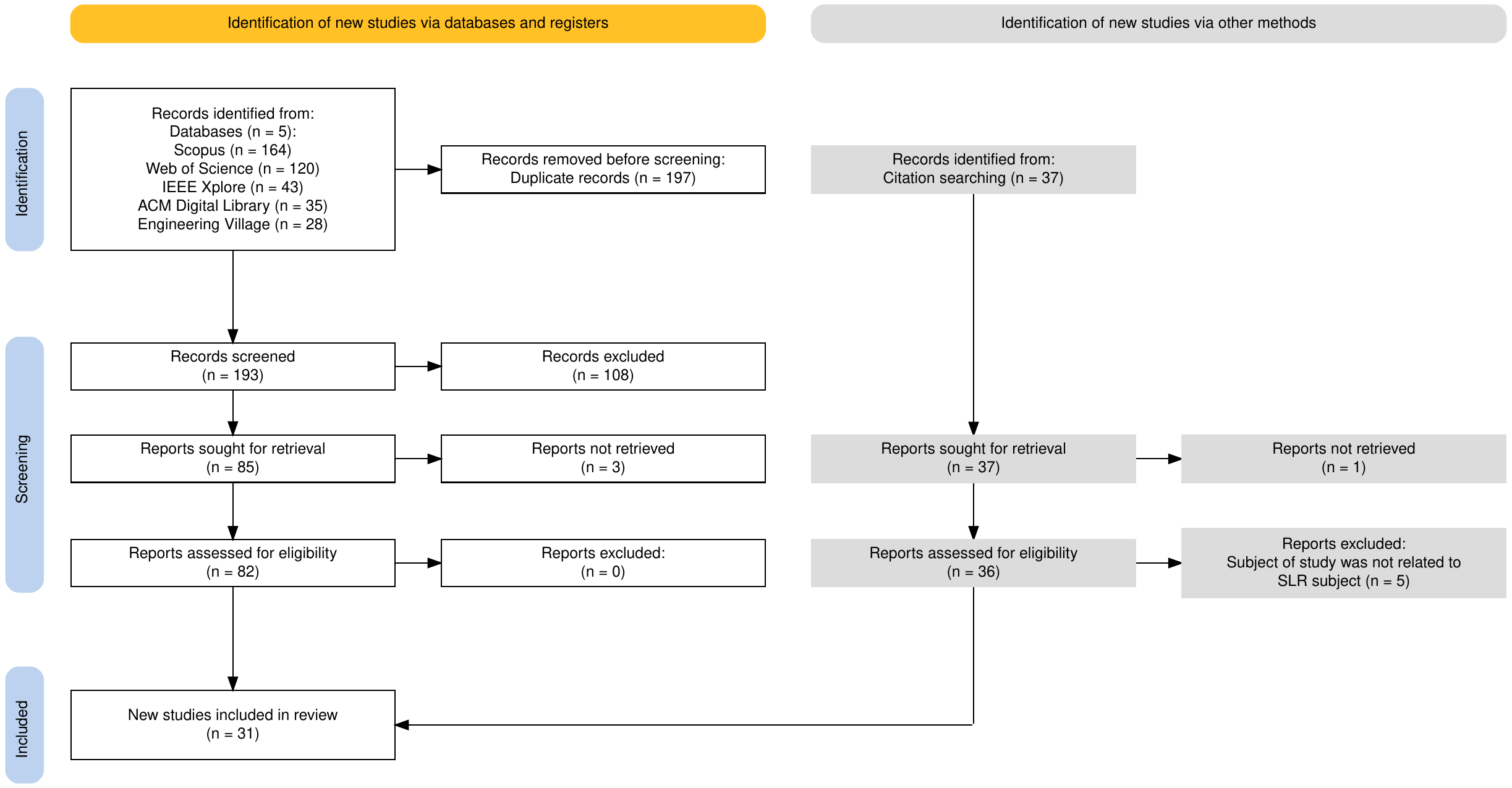}
    \caption{PRISMA 2020 flow diagram for the SLR}
    \label{fig:prisma}
\end{figure}

No suitable references (except for supporting information) predated 2021. The resulting corpus consisted of 115 journal articles, 85 papers in conference proceedings, and 11 book sections.
Table~\ref{tab:publicationsPerSource} shows a breakdown of papers per database and library. Figure \ref{fig:publicationsPerYear} shows the near doubling of publication volume in the last two years (2023-2024).
The PRISMA flow diagram \cite{haddaway2022prismaApp2020} in Figure \ref{fig:prisma} documents the screening and selection process.


\subsection{Quality Assessment}
\label{sec:qa}
Following the PRISMA guidelines, each study was evaluated based on: clarity of research questions and objectives; appropriateness of research methodology; relevance of findings to digital twins, systems-of-systems, verification and/or validation of digital twins; credibility of data sources and study reliability. After full-text reading and assessment of papers, the final selection was 21 papers, published between 2022 and 2024, that discuss a DT composition approach and address a form of verification and/or validation of their DT structure.

\subsection{Data Extraction and Synthesis}
Data from selected studies were systematically extracted and synthesized based on:
study metadata (authors, year); DT application domains; composition approach used; DT properties \& qualities evaluated; V\&V techniques used; formal methods or frameworks employed; V\&V challenges and gaps identified. 

\section{Results \& Discussion}
\label{sec:results&disscussion}
In this section, we detail and comment on the findings of our systematic review, organised by the research questions posed in Section \ref{sec:objective}. We discuss the identified DT composition approaches, and then consider DT properties and qualities, current V\&V methods, and recurring challenges in the field.

\subsection{RQ1: Composition Approaches for Digital Twins}
\begin{figure}[htbp]
    \centering
    \includegraphics[width=0.6\linewidth]{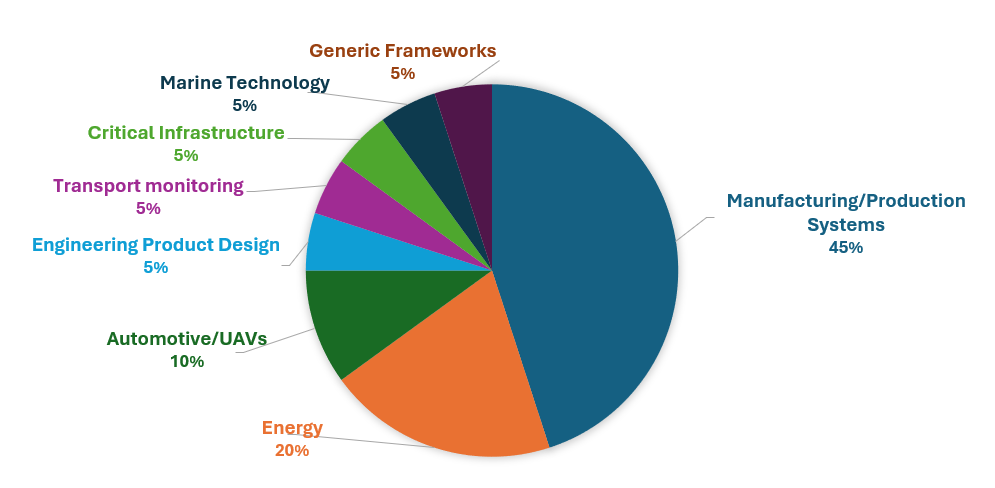}
    \caption{Application Domain Distribution within SoS}
    \label{fig:App_Domains}
\end{figure}

Our analysis identified seven application domains in which DT composition approaches have been explored (Figure \ref{fig:App_Domains}). Manufacturing and production systems represent the most prominent domain with nine papers, followed by energy applications with four papers. This distribution reflects the industrial focus of DT composition research, which emphasises complex production environments.

We identified seven approaches to DT composition in the reviewed papers (Table \ref{tab:composition-approaches}).  Note that, here and in the subsequent tables, we use the various terms and descriptions found in the literature, but would not at this stage endorse these particular definitions. Orchestrated integration was the dominant approach, employed in 11 papers, where a centralized mechanism manages interactions between DT components. Other approaches include federated integration, service-based integration, and co-simulation, each addressing specific integration needs in different application contexts.

\begin{table}[ht]
\centering
\caption{Classification of DT Composition Approaches}
\label{tab:composition-approaches}
\begin{tabular}{@{}
    >{\raggedright\arraybackslash}p{2.2cm}
    >{\raggedright\arraybackslash}p{7.8cm}
    >{\raggedright\arraybackslash}p{2cm}
@{}} \hline
\textbf{Approach} & \textbf{Description} & \textbf{Papers} \\ \hline
Orchestrated Integration & DTs are composed through a centralized orchestrator that manages the interaction, data flow, and behaviour between DT components & \cite{Leng2023DTPlatformReconfigMS,Overbeck2024DTProdSys,He2024DTSituationAware,Dahmen2022V&VDTVirtTestbeds} 
\cite{Temperekidis2023DTFormalAnalys,Wu2022DTmultidispcollabDes,VanMark2021ModelMeetData,Wen2024DecompDT} 
\cite{Gill2024Pipeline,Liu2023DTShopFloor,Barbone2024DTContinuum} \\ \hline
Federated Integration & Multiple autonomous DTs collaborate through decentralized aggregation mechanisms, while maintaining individual operation & \cite{Aloqaily2023Ind4DTBlockchainFL,Zhou2024FedDTUAV} \\ \hline
Service-based Integration & DTs are composed as loosely coupled services that interact through defined interfaces & \cite{Park2023DTFrameRCB,Perisic2024Helix} \\ \hline
Co-simulation & DTs are composed through simulation interfaces that enable information exchange between different models & \cite{Jeon2024ReliabFrameDT} \\ \hline
Multi-model Integration & Multiple models are run concurrently and their outputs combined for improved performance & \cite{Liu2024MultiDT} \\ \hline
Decentralized Coordination & Distributed DTs coordinate through communication protocols without central control & \cite{Samadi2024MicrogridDT} \\ \hline
Hierarchical Integration & DTs are organized and composed in a hierarchical structure reflecting the physical system's organization & \cite{Dobaj2022DTenabDevOPs,Novak2022DigiAutoEngInd4} \\ \hline
\end{tabular}
\end{table}

The most frequently addressed SoS characteristics in DT composition were emergence, heterogeneity, distributed nature, managerial independence, and complexity. These characteristics significantly influence the choice of composition approach: emergence typically drives orchestrated approaches for coordinating system-level behaviours \cite{He2024DTSituationAware,Overbeck2024DTProdSys}; heterogeneity leads to service-based integration patterns with standardized interfaces \cite{Park2023DTFrameRCB,Perisic2024Helix}; distributed nature corresponds with federated approaches maintaining component autonomy \cite{Aloqaily2023Ind4DTBlockchainFL,Zhou2024FedDTUAV}; while complexity generally requires sophisticated orchestrated architectures with hierarchical structures \cite{Gordan2024PRECINCT,Liu2023DTShopFloor}. The most comprehensive solutions addressing multiple SoS characteristics employ hybrid patterns that combine elements of orchestration with service-based or federated mechanisms \cite{Gill2024Pipeline,Gordan2024PRECINCT}.

\subsection{RQ2: Digital Twin Properties and Qualities}

By examining system-level properties and quality attributes of DTs addressed in V\&V studies, we aim to identify which properties and qualities researchers prioritize when developing V\&V approaches for DTs, revealing both established patterns and potential gaps in the literature.

\begin{table}[ht]
\centering
\caption{Properties of Digital Twins}
\label{tab:DT-properties}
\begin{tabular}{@{}
    >{\raggedright\arraybackslash}p{2cm}
    >{\raggedright\arraybackslash}p{2cm}
    >{\raggedright\arraybackslash}p{6.4cm}
    >{\raggedright\arraybackslash}p{1.5cm}
@{}} \hline
\textbf{Category} & \textbf{Property} & \textbf{Description} & \textbf{Papers} \\ \hline
Fidelity & Accuracy & Degree to which the DT correctly represents the physical entity & \cite{Overbeck2024DTProdSys,Park2023DTFrameRCB,Liu2024MultiDT} \\ \cline{2-4}
 & Twinning Fidelity & How accurately the DT mirrors its physical counterpart, sometimes adaptive across different levels & \cite{Dobaj2022DTenabDevOPs,Jeon2024ReliabFrameDT,Perisic2024Helix} \\ \cline{2-4}
 & Represent-ability & Ability to properly represent the physical entity's characteristics & \cite{He2024DTSituationAware} \\ \hline
Temporal & Real-time Synch. & Ability to maintain consistency between physical and digital entities in real-time & \cite{Leng2023DTPlatformReconfigMS,Liu2023DTShopFloor} \\ \cline{2-4}
 & Latency & Delay between changes in the PT and corresponding updates in the DT & \cite{Dobaj2022DTenabDevOPs,Wu2022DTmultidispcollabDes,Barbone2024DTContinuum,Samadi2024MicrogridDT} \\ \cline{2-4}
 & Interactivity & Ability to interact with PT in real-time & \cite{He2024DTSituationAware} \\ \hline
Structural & Interoper-ability & Ability of different DT components to work together seamlessly & \cite{He2024DTSituationAware,Park2023DTFrameRCB} \cite{Temperekidis2023DTFormalAnalys,Gill2024Pipeline,Barbone2024DTContinuum} \\ \cline{2-4}
 & Consistency & Maintaining coherent states across models or between physical and digital entities & \cite{VanMark2021ModelMeetData,Liu2023DTShopFloor,Gill2024Pipeline} \\ \cline{2-4}
 & Modularity & Degree to which DT components can be separated and recombined & \cite{Wen2024DecompDT} \\ \cline{2-4}
 & Scalability & Ability to handle increasing amounts of data or components & \cite{Barbone2024DTContinuum,Samadi2024MicrogridDT} \\ \hline
Behavioural & Predictability & Ability to forecast system behaviour & \cite{He2024DTSituationAware} \\ \cline{2-4}
 & Subjectivity & Ability to provide personalized views & \cite{He2024DTSituationAware} \\ \cline{2-4}
 & Self-Evolution & Capacity to adapt and learn autonomously & \cite{He2024DTSituationAware} \\ \cline{2-4}
 & Adaptability & Ability to adjust to changing system configurations or environments & \cite{Dobaj2022DTenabDevOPs,Liu2024MultiDT} \\ \hline
Trust-worthiness & Correctness & Conformity to specified requirements or standards & \cite{Dahmen2022V&VDTVirtTestbeds,Gill2024Pipeline} \\ \cline{2-4}
 & Numerical Stability & Stability of numerical calculations within the DT models & \cite{Dahmen2022V&VDTVirtTestbeds} \\ \cline{2-4}
 & Integrity & Maintenance of system integrity in operation & \cite{Dobaj2022DTenabDevOPs} \\ \cline{2-4}
 & Availability & Readiness for correct service & \cite{Dobaj2022DTenabDevOPs} \\ \hline
\end{tabular}
\end{table}

Our analysis identified several key DT properties (fundamental characteristics essential to DT functionality) addressed in the literature, which we grouped as shown in Table \ref{tab:DT-properties}. Structural properties, particularly interoperability, were most frequently addressed, highlighting the importance of integration capabilities in composed DT systems. Fidelity and temporal properties were also prominent, focusing on the accuracy of representation and real-time performance requirements.

We identified six categories of quality attributes (non-functional aspects affecting DT performance and usability) addressed in the literature (Table \ref{tab:DT-qualities}). Adaptability qualities were most frequently discussed, reflecting the need for DTs to respond to changing conditions in complex systems. Verification qualities were also prominent, emphasizing the importance of validating DT behaviour against real-world counterparts in composed systems.

\begin{table}[ht]
\centering
\caption{Qualities of Digital Twins}
\label{tab:DT-qualities}
\begin{tabular}{@{}
    >{\raggedright\arraybackslash}p{2.0cm}
    >{\raggedright\arraybackslash}p{2.3cm}
    >{\raggedright\arraybackslash}p{6.0cm} 
    >{\raggedright\arraybackslash}p{1.6cm}   
@{}} \hline
\textbf{Category} & \textbf{Quality} & \textbf{Description} & \textbf{Papers} \\ \hline
Verification & Verifiability & Ability to confirm that the DT accurately represents the physical entity & \cite{Dahmen2022V&VDTVirtTestbeds,Perisic2024Helix,Temperekidis2023DTFormalAnalys,Wu2022DTmultidispcollabDes} \\ \cline{2-4}
 & Reliability & Ability to perform correctly under stated conditions for a specified period & \cite{Novak2022DigiAutoEngInd4,Jeon2024ReliabFrameDT} \\ \cline{2-4}
 & Validity & Degree to which the DT remains relevant over the system lifecycle & \cite{Overbeck2024DTProdSys} \\ \hline
Adaptability & Self-adaptation & Ability to adapt autonomously to changing conditions & \cite{Dobaj2022DTenabDevOPs} \\ \cline{2-4}
 & Continuous optimization & Ongoing improvement of performance and efficiency & \cite{Dobaj2022DTenabDevOPs} \\ \cline{2-4}
 & Flexibility & Ability to accommodate changes in the physical system & \cite{Novak2022DigiAutoEngInd4,Overbeck2024DTProdSys} \\ \cline{2-4}
 & Resilience & Ability to maintain acceptable operation in the face of disturbances & \cite{Dobaj2022DTenabDevOPs,Novak2022DigiAutoEngInd4,Park2023DTFrameRCB,Gordan2024PRECINCT} \\ \hline
Interoper-ability & Vert./Horiz. interoperability & Ability to integrate with systems at different levels and domains & \cite{Dobaj2022DTenabDevOPs,Novak2022DigiAutoEngInd4} \\ \cline{2-4}
 & Platform independence & Ability to operate across different computing platforms & \cite{Barbone2024DTContinuum} \\ \cline{2-4}
 & Semantic interoperability & Ability to exchange information meaningfully between different systems & \cite{Wen2024DecompDT} \\ \hline
Management & Maintainability & Ease with which DT can be maintained & \cite{Overbeck2024DTProdSys,Wen2024DecompDT} \\ \cline{2-4}
 & Efficiency & Optimal use of resources & \cite{Gill2024Pipeline,Wen2024DecompDT,Barbone2024DTContinuum} \\ \cline{2-4}
 & Reusability & Ability to reuse components & \cite{Jeon2024ReliabFrameDT,Wen2024DecompDT} \\ \hline
Awareness & Situation awareness & Ability to perceive environmental elements within time and space & \cite{He2024DTSituationAware} \\ \cline{2-4}
 & Network awareness & Consideration of network conditions in DT operation & \cite{Barbone2024DTContinuum} \\ \hline
Trust & Privacy & Protection of sensitive data & \cite{Aloqaily2023Ind4DTBlockchainFL} \\ \cline{2-4}
 & Trustworthiness & Degree to which DT can be relied upon & \cite{Zhou2024FedDTUAV,Gordan2024PRECINCT} \\ \cline{2-4}
 & Autonomy & Ability to operate independently & \cite{Zhou2024FedDTUAV} \\ \hline
\end{tabular}
\end{table}

\subsection{RQ3: Verification and Validation Approaches for Digital Twins}

Based on the analysis of the reviewed papers, V\&V approaches for digital twins can be classified into three levels of formality: formal, semi-formal, and informal. Table \ref{tab:V&V-Approaches} presents this classification with the corresponding approaches, descriptions, and references to the papers that utilized these approaches.

\begin{table}[ht] 
\centering
\caption{V\&V Approaches Based on Formality Level}
\label{tab:V&V-Approaches}
\begin{tabular}{@{}
    >{\raggedright\arraybackslash}p{1.3cm}
    >{\raggedright\arraybackslash}p{2.2cm}
    >{\raggedright\arraybackslash}p{6.5cm}
    >{\raggedright\arraybackslash}p{1.9cm}
@{}} \hline
\textbf{Level} & \textbf{Approach} & \textbf{Description} & \textbf{Papers} \\ \hline
Formal & Model Checking & Systematic verification against formal specifications using temporal logic & \cite{Temperekidis2023DTFormalAnalys,VanMark2021ModelMeetData,Dahmen2022V&VDTVirtTestbeds} \\ \cline{2-4}
 & Theorem Proving & Mathematical verification of system properties using axioms and inference rules & \cite{VanMark2021ModelMeetData,Gill2024Pipeline} \\ \cline{2-4}
 & Semantic Validation & Validation of models against domain-specific rules using semantic web technologies & \cite{Gill2024Pipeline,Novak2022DigiAutoEngInd4} \\ \hline
Semi-formal & Simulation-Based Testing & Validation through simulated scenarios that represent real-world conditions & \cite{Leng2023DTPlatformReconfigMS,Overbeck2024DTProdSys,Dobaj2022DTenabDevOPs,Wu2022DTmultidispcollabDes} \cite{Samadi2024MicrogridDT,Zhou2024FedDTUAV,Gordan2024PRECINCT,Jeon2024ReliabFrameDT} \\ \cline{2-4}
 & Co-simulation & Integration of multiple simulation models and tools to validate system behaviour & \cite{Temperekidis2023DTFormalAnalys,Wu2022DTmultidispcollabDes,Dahmen2022V&VDTVirtTestbeds} \\ \cline{2-4}
 & Runtime Verification & Monitoring system behaviour during execution to verify compliance with specifications & \cite{Temperekidis2023DTFormalAnalys,Dobaj2022DTenabDevOPs} \\ \cline{2-4}
 & Model-Based Testing & Using models as the basis for test case generation and evaluation & \cite{Perisic2024Helix,Wu2022DTmultidispcollabDes,Dahmen2022V&VDTVirtTestbeds} \\ \cline{2-4}
 & DT Continuum Orchestration & Coordinating and monitoring DT instances across platforms & \cite{Barbone2024DTContinuum} \\ \hline
Informal & Experimental Validation & Testing in real-world environments to validate DT performance & \cite{Leng2023DTPlatformReconfigMS,Novak2022DigiAutoEngInd4,Park2023DTFrameRCB,Zhou2024FedDTUAV} \\ \cline{2-4}
 & Case Studies & Demonstrating DT application & \cite{He2024DTSituationAware,Liu2023DTShopFloor,Wen2024DecompDT,Liu2024MultiDT} \\ \cline{2-4}
 & Empirical Metrics & Using quantitative performance metrics to assess DT accuracy & \cite{Overbeck2024DTProdSys,Park2023DTFrameRCB,Liu2024MultiDT,Aloqaily2023Ind4DTBlockchainFL} \\ \cline{2-4}
 & Expert Review & Manual evaluation by domain experts & \cite{He2024DTSituationAware,Gordan2024PRECINCT} \\ \hline
\end{tabular}
\end{table}

While formal approaches provide mathematical rigour and stronger guarantees, they often face scalability challenges when applied to complex DT systems. Semi-formal approaches, particularly simulation-based testing and co-simulation, represent the most widely adopted methods due to their balance of structure and flexibility. Informal approaches such as experimental validation and case studies provide practical insights but offer limited guarantees.

The scope of V\&V for DTs encompasses multiple dimensions. The primary focus areas in the reviewed literature are listed below.
\begin{enumerate}
    \item \textbf{Model Fidelity Validation,} mentioned in 11 papers (52.4\%), focuses on ensuring the DT accurately represents the structure, components, and parameters of the PT~\cite{Leng2023DTPlatformReconfigMS,Overbeck2024DTProdSys,Dahmen2022V&VDTVirtTestbeds}.
    \item \textbf{Behavioural Correctness Verification,} addressed in 14 papers (66.7\%), concerns ensuring that the DT properly simulates and predicts physical system behaviour \cite{Temperekidis2023DTFormalAnalys,Wu2022DTmultidispcollabDes,Samadi2024MicrogridDT}.
    \item \textbf{Integration Validation,} featured in 9 papers (42.9\%), concerns interactions between components and systems \cite{Gill2024Pipeline,Jeon2024ReliabFrameDT,Barbone2024DTContinuum}.
    \item \textbf{Performance Assessment,} present in 8 papers (38.1\%), concerns the evaluation of DT operational metrics such as accuracy, response time, and robustness \cite{Park2023DTFrameRCB,Liu2024MultiDT,Zhou2024FedDTUAV}.
    \item \textbf{Cyber-Physical Consistency,} emphasized in 7 papers (33.3\%), focuses on maintaining alignment between physical and digital entities \cite{Liu2023DTShopFloor,Dobaj2022DTenabDevOPs,Overbeck2024DTProdSys}.
\end{enumerate}

The data indicate an evolution in V\&V scope emphasis. Earlier papers (2022-2023) focused more on basic model validation, while recent papers (2023-2024) show increased attention to integration challenges and multi-domain verification. Runtime verification and cyber-physical consistency have gained prominence in 2023-2024 publications. This suggests that, while behavioural correctness verification is the predominant V\&V focus, there is a growing emphasis on integration validation and cyber-physical consistency as DT implementations become more complex and interconnected.

\subsection{RQ4: Challenges in V\&V of Digital Twins}

The systematic review identified several challenges related to verification and validation of digital twins. Table \ref{tab:V&V-challenges} categorizes these challenges along with descriptions and the corresponding papers that mentioned them.

\begin{table}[ht]
\centering
\caption{Classification of V\&V Challenges}
\label{tab:V&V-challenges}
\begin{tabular}{@{}
    >{\raggedright\arraybackslash}p{1.6cm}
    >{\raggedright\arraybackslash}p{2.8cm}
    >{\raggedright\arraybackslash}p{5.9cm}
    >{\raggedright\arraybackslash}p{1.6cm}
@{}} \hline
\textbf{Type} & \textbf{Challenge} & \textbf{Description} & \textbf{Papers} \\ \hline
Technical & Model Uncertainty & Difficulties in quantifying and managing uncertainties in DT models & \cite{VanMark2021ModelMeetData,Perisic2024Helix,Gordan2024PRECINCT} \\ \cline{2-4}
 & Real-time Synchronization & Maintaining accurate alignment between PT and DT in real-time & \cite{Dobaj2022DTenabDevOPs,Zhou2024FedDTUAV,Samadi2024MicrogridDT} \\ \cline{2-4}
 & Scalability & Challenges in scaling V\&V approaches to large-scale DT systems & \cite{VanMark2021ModelMeetData,Dahmen2022V&VDTVirtTestbeds,Barbone2024DTContinuum,Samadi2024MicrogridDT} \\ \cline{2-4}
 & Integration Complexity & Difficulties in V\&V of integrated heterogeneous models and components & \cite{Jeon2024ReliabFrameDT,Gill2024Pipeline,Barbone2024DTContinuum,Temperekidis2023DTFormalAnalys} \\ \cline{2-4}
 & Dynamic System Changes & Handling structural or behavioural change in PT to be reflected in DT & \cite{Overbeck2024DTProdSys,Dobaj2022DTenabDevOPs,Zhou2024FedDTUAV} \\ \hline
Methodo-logical & Standardized Methods & Lack of established standards and methodologies for DT V\&V & \cite{Perisic2024Helix,Wen2024DecompDT,Jeon2024ReliabFrameDT} \\ \cline{2-4}
 & Multi-domain Verification & Difficulties in verifying DTs that span multiple engineering domains & \cite{Wen2024DecompDT,Gill2024Pipeline,Gordan2024PRECINCT} \\ \cline{2-4}
 & Formal Verif. Complexity & Challenges in applying formal methods to complex DT systems & \cite{VanMark2021ModelMeetData,Temperekidis2023DTFormalAnalys,Dahmen2022V&VDTVirtTestbeds} \\ \cline{2-4}
 & Metrics & Identifying suitable metrics for assessing DT fidelity and performance & \cite{Overbeck2024DTProdSys,Temperekidis2023DTFormalAnalys,Dahmen2022V&VDTVirtTestbeds} \\ \hline
Data-related & Data Quality and Availability & Issues with obtaining sufficient high-quality data for validation & \cite{Gordan2024PRECINCT,Zhou2024FedDTUAV,Samadi2024MicrogridDT} \\ \cline{2-4}
 & Model-Data Integration & Challenges in integrating heterogeneous data sources with models & \cite{Barbone2024DTContinuum,He2024DTSituationAware,Zhou2024FedDTUAV} \\ \hline
Practical & Resource Constraints & Limited computational, time, or expertise resources for comprehensive V\&V & \cite{Dobaj2022DTenabDevOPs,Barbone2024DTContinuum,Samadi2024MicrogridDT} \\ \cline{2-4}
 & Tool Interoperability & Integration challenges between different V\&V tools and platforms & \cite{Perisic2024Helix,Jeon2024ReliabFrameDT,Barbone2024DTContinuum} \\ \hline
\end{tabular}
\end{table}

The identified challenges highlight several important research directions for the field, including the development of standardized V\&V methodologies specifically designed for digital twins, techniques for handling model uncertainty and dynamic system changes, and approaches for efficient integration and validation of heterogeneous models. Addressing these challenges will be critical for ensuring the reliability and trustworthiness of digital twins as they continue to be deployed in increasingly complex and safety-critical domains. Our experience in this SLR suggests that it will be important to clarify common definitions for features and properties such as DT fidelity in order to facilitate coherent research on V\&V.

\section{Threats to Validity}
\label{sec:threats}

Our study faces a number of validity threats. Construct validity is challenged by inconsistent terminology (e.g., Digital Twin vs. Virtual Twin). We treated these as distinct unless a paper clearly used them interchangeably. Also, papers with Titles using ``Twin'' only without ``Digital'' have not appeared in the search results. This may have excluded relevant work using different terms.

Initial screening used only titles, abstracts, and keywords, adding to the risk of missing some relevant papers. To reduce this, uncertain cases were included for full review. Classification also posed challenges, as studies often span dimensions. We used established schemes and allowed overlaps when needed.

External validity is limited by focusing on peer-reviewed, English, online-accessible papers. This ensures quality but reduces generalizability. So-called ``grey literature'' and Google Scholar were not used, due to the relative lack of filtering and transparency.

Internal validity is affected by the possibility of selection bias due to limiting the search to the CPS domain. Meanwhile, studies discussing V\&V of DTs from a general perspective without explicitly mentioning CPSs would not appear in the corpus except if found in the manual search. Team discussions and early paper reviews helped align our mapping. Finally, a few promising papers were excluded because their full text is unavailable, even though their abstract suggests they fit the criteria. Full methods are documented for transparency and reproducibility.

\section{Conclusion}
\label{sec:conclusion}
Our goal in this SLR is to support research and innovation in V\&V in the engineering of DTs for SoSs. The review has identified growing interest in DT integration in SoS contexts, but highlighted that formal approaches to composition and rigorous V\&V remain underdeveloped. Most studies prioritize simulation-based validation or qualitative assessments, with limited attention to continuous or scalable V\&V frameworks. Future research should focus on formal support for DT composition and establishing standardized, domain-independent V\&V methods to ensure trustworthy and interoperable DT deployments in SoS.

Our future direction is to examine the INTO-CPS \href{https://github.com/INTO-CPS-Association/plant-controller/tree/v0.1}{``greenhouse'' case study} for its compositionality requirements and SoS-level properties. This project will be built from the perspective of model-based SoS engineering to allow us to examine the creation of DTs in such a setting. We see this as a starting point to discovering V\&V methods and tools that can be applied to DT composition more widely.


\noindent
\\
\small
\textbf{\ackname} The authors are grateful to Newcastle University for access to online resources for the SLR reported here. The authors are grateful to anonymous reviewers for their valuable comments on the first draft of this paper.

\noindent\textbf{\discintname} The authors have no competing interests to declare.

\bibliographystyle{splncs04}
\bibliography{references}

\begin{thebibliography}{10}
\providecommand{\url}[1]{\texttt{#1}}
\providecommand{\urlprefix}{URL }
\providecommand{\doi}[1]{https://doi.org/#1}

\bibitem{Aloqaily2023Ind4DTBlockchainFL}
Aloqaily, M., Ridhawi, I.A., Kanhere, S.: Reinforcing industry 4.0 with digital twins and blockchain-assisted federated learning. IEEE Journal on Selected Areas in Communications  \textbf{41}(11),  3504--3516 (2023)

\bibitem{Barbone2024DTContinuum}
Barbone, A., Burattini, S., Martinelli, M., Picone, M., Ricci, A., Virdis, A.: Digital twin continuum: a key enabler for pervasive cyber-physical environments (2024). \doi{10.1109/icccn61486.2024.10637565}

\bibitem{Bitencourt2025V&VSLR}
Bitencourt, J., Wooley, A., Harris, G.: Verification and validation of digital twins: a systematic literature review for manufacturing applications. Intl. Jnl. of Production Research  \textbf{63}(1),  342--370 (2024)

\bibitem{Borth2019}
Borth, M., Verriet, J., Muller, G.: {Digital Twin Strategies for SoS}. In: 14th Annual Conf. System of Systems Engineering. pp. 164--169 (2019). \doi{10.1109/SYSOSE.2019.8753860}

\bibitem{BOYES2022103763}
Boyes, H., Watson, T.: Digital twins: An analysis framework and open issues. Computers in Industry  \textbf{143},  103763 (2022). \doi{10.1016/j.compind.2022.103763}

\bibitem{EvertonCavalante2024}
Cavalcante, E., Batista, T.: Exploring synergies and challenges of digital twins in systems-of-systems (2024), manuscript submitted to 1st Intl. Conf. on Engineering Digital Twins. ACM, New York, NY, USA

\bibitem{Dahmen2022V&VDTVirtTestbeds}
Dahmen, U., Osterloh, T., Roßmann, J.: Veriﬁcation and validation of digital twins and virtual testbeds. Intl. Jnl. of Advances in Applied Sciences  \textbf{11}(1) (2022)

\bibitem{Diakité2023}
Diakité, M., Traoré, M.K.: Formal approach to digital twin specification. In: 2023 Annual Modeling and Simulation Conference. pp. 233--244 (2023)

\bibitem{Dobaj2022DTenabDevOPs}
Dobaj, J., Riel, A., Krug, T., et~al.: {Towards Digital Twin-enabled DevOps for CPS providing Architecture-Based Service Adaptation \& Verification at Runtime}. In: 17th Intl. Symp. on Software Engineering for Adaptive and Self-Managing Systems. pp. 132--143. IEEE (2022). \doi{10.1145/3524844.3528057}

\bibitem{EstGomm2021}
Esterle, L., Gomes, C., Frasheri, M., et~al.: Digital twins for collaboration and self-integration. In: Intl. Conf. on Autonomic Computing and Self-Organizing Systems Companion. pp. 172--177 (2021). \doi{10.1109/ACSOS-C52956.2021.00040}

\bibitem{Fett2023}
Fett, M., Wilking, F., Goetz, S., et~al.: A literature review on the development and creation of digital twins, cyber-physical systems, and product-service systems. Sensors  \textbf{23}(24) (2023). \doi{10.3390/s23249786}

\bibitem{Fitzgerald2024}
Fitzgerald, J., Gomes, C., Larsen, P.G. (eds.): {The Engineering of Digital Twins}. Springer (2024)

\bibitem{Fitzgerald2019}
Fitzgerald, J., Larsen, P.G., Pierce, K.: Multi-modelling and Co-simulation in the Engineering of Cyber-Physical Systems: Towards the Digital Twin, pp. 40--55. Springer, Cham (2019)

\bibitem{Gill2024Pipeline}
Gill, M.S., Zhang, J., Wortmann, A., Fay, A.: Toward automating the composition of digital twins within system-of-systems. In: 29th IEEE Intl. Conf. on Emerging Technologies and Factory Automation. pp.~1--4 (2024). \doi{10.1109/ETFA61755.2024.10710740}

\bibitem{Gordan2024PRECINCT}
Gordan, M., Kountche, D.A., McCrum, D., et~al.: Protecting critical infrastructure against cascading effects: The precinct approach. Resilient Cities and Structures  \textbf{3}(3),  1--19 (2024). \doi{10.1016/j.rcns.2024.04.001}

\bibitem{haddaway2022prismaApp2020}
Haddaway, N.R., Page, M.J., Pritchard, C.C., McGuinness, L.A.: {PRISMA2020: An R package and Shiny app for producing PRISMA 2020-compliant flow diagrams, with interactivity for optimised digital transparency and Open Synthesis}. Campbell Systematic Reviews  \textbf{18}(2),  e1230 (2022)

\bibitem{He2024DTSituationAware}
He, X., Tang, Y., Ma, S., et~al.: Redefinition of digital twin and its situation awareness framework designing toward fourth paradigm for energy internet of things. IEEE Transactions on Systems, Man, and Cybernetics: Systems  \textbf{54}(11),  6873--6888 (2024). \doi{10.1109/TSMC.2024.3407061}

\bibitem{incose2023handbook}
INCOSE: INCOSE Systems Engineering Handbook. Wiley (2023)

\bibitem{Jeon2024ReliabFrameDT}
Jeon, H., Nguyen, D.T., Zadeh, M.: Reliability-based design framework for assessment of digital twin models applied to vessel controllers (2024). \doi{10.1109/ecce55643.2024.10861242}

\bibitem{Kalyvas2021}
Kalyvas, M.: An innovative industrial control system architecture for real-time response, fault-tolerant operation and seamless plant integration. The Journal of Engineering  \textbf{2021}(10),  569--581 (2021). \doi{10.1049/tje2.12064}

\bibitem{Kitchenham2004SLR}
Kitchenham, B.: Procedures for performing systematic reviews. Keele, UK, Keele Univ.  \textbf{33} (08 2004)

\bibitem{Lana2019}
Lana, C.A., Guessi, M., Antonino, P.O., et~al.: A systematic identification of formal and semi-formal languages and techniques for software-intensive systems-of-systems requirements modeling. IEEE Systems Journal  \textbf{13}(3),  2201--2212 (2019). \doi{10.1109/JSYST.2018.2874061}

\bibitem{Leng2023DTPlatformReconfigMS}
Leng, B.H., Gao, S., Xia, T.B., et~al.: Digital twin monitoring and simulation integrated platform for reconfigurable manufacturing systems. Advanced Engineering Informatics  \textbf{58}, ~14 (2023). \doi{10.1016/j.aei.2023.102141}

\bibitem{Liu2023DTShopFloor}
Liu, H., Zhang, J., Cheng, Y., et~al.: Digital twin shop-floor: A complex system-oriented construction method and operation mechanism. In: Procedia CIRP. vol.~119, pp. 52--57. Elsevier (2023). \doi{10.1016/j.procir.2023.03.083}

\bibitem{Liu2024MultiDT}
Liu, S., Tian, J., Ji, Z., et~al.: Research on multi-digital twin and its application in wind power forecasting. Energy  \textbf{292} (2024). \doi{10.1016/j.energy.2024.130269}

\bibitem{lv2022DTsBook}
Lv, Z., Fersman, E.: Digital twins: basics and applications. Springer (2022)

\bibitem{Mandel2022}
Mandel, C., Guenther, M., Martin, A., et~al.: Towards a system of systems engineering architecture framework. In: 17th Annual System of Systems Engineering Conf. pp. 221--226 (2022). \doi{10.1109/SOSE55472.2022.9812634}

\bibitem{JudithMichael2022}
Michael, J., Pfeiffer, J., Rumpe, B., Wortmann, A.: Integration challenges for digital twin systems-of-systems. In: IEEE/ACM 10th Intl. Workshop on Software Engineering for Systems-of-Systems and Software Ecosystems. pp. 9--12 (2022)

\bibitem{Moyne2020}
Moyne, J., Qamsane, Y., Balta, E.C., et~al.: A requirements driven digital twin framework: Specification and opportunities. IEEE Access  \textbf{8},  107781--107801 (2020). \doi{10.1109/ACCESS.2020.3000437}

\bibitem{Nguyen2019}
Nguyen, T.: {Formal Requirements and Constraints Modelling in FORM-L for the Engineering of Complex Socio-Technical Systems}. In: IEEE 27th Intl. Requirements Engineering Conference Workshops. pp. 123--132 (2019). \doi{10.1109/REW.2019.00027}

\bibitem{Novak2022DigiAutoEngInd4}
Novák, P., Vyskočil, J.: {Digitalized Automation Engineering of Industry 4.0 Production Systems and Their Tight Cooperation with Digital Twins}. Processes  \textbf{10}(2), ~27 (2022)

\bibitem{Olsson2023}
Olsson, T., Axelsson, J.: Systems-of-systems and digital twins: A survey and analysis of the current knowledge. In: 2023 18th Annual System of Systems Engineering Conf. pp.~1--6 (2023). \doi{10.1109/SoSE59841.2023.10178527}

\bibitem{Overbeck2024DTProdSys}
Overbeck, L., Graves, S.C., Lanza, G.: Development and analysis of digital twins of production systems. Intl. Jnl. of Production Research  \textbf{62}(10),  3544--3558 (2024)

\bibitem{Pagen71PRISMA}
Page, M.J., McKenzie, J.E., Bossuyt, P.M., et~al.: {The PRISMA 2020 statement: an updated guideline for reporting systematic reviews}. BMJ  \textbf{372} (2021). \doi{10.1136/bmj.n71}

\bibitem{Park2023DTFrameRCB}
Park, K.T., Park, Y.H., Park, M.W., Noh, S.D.: Architectural framework of digital twin-based cyber-physical production system for resilient rechargeable battery production. Journal of Computational Design and Engineering  \textbf{10}(2),  809--829 (2023)

\bibitem{Perisic2024Helix}
Perisic, A., Perisic, B.: Digital twins verification and validation approach through the quintuple helix conceptual framework. Electronics  \textbf{13}(16), ~32 (2024). \doi{10.3390/electronics13163303}

\bibitem{Piras2024}
Piras, G., Agostinelli, S., Muzi, F.: Digital twin framework for built environment: A review of key enablers. Energies  \textbf{17}(2) (2024). \doi{10.3390/en17020436}

\bibitem{Samadi2024MicrogridDT}
Samadi, M., Fattahi, J.: Low-inertia microgrid synchronization using data-driven digital twins. IEEE Access  \textbf{12},  78534--78548 (2024). \doi{10.1109/ACCESS.2024.3408715}

\bibitem{Samak2023}
Samak, T., Samak, C., Kandhasamy, S., Krovi, V., Xie, M.: {AutoDRIVE: A Comprehensive, Flexible and Integrated Digital Twin Ecosystem for Autonomous Driving Research \& Education}. Robotics  \textbf{12}(3) (2023). \doi{10.3390/robotics12030077}

\bibitem{Seo2016}
Seo, D., Shin, D., Baek, Y.M., et~al.: Modeling and verification for different types of system of systems using prism. In: 2016 IEEE/ACM 4th International Workshop on Software Engineering for Systems-of-Systems (SESoS). pp. 12--18 (2016). \doi{10.1145/2897829.2897833}

\bibitem{Temperekidis2023DTFormalAnalys}
Temperekidis, A., Kekatos, N., Katsaros, P., et~al.: Towards a digital twin architecture with formal analysis capabilities for learning-enabled autonomous systems. In: Lecture Notes in Computer Science. vol. 13866, pp. 163--181. Springer (2023). \doi{10.1007/978-3-031-31268-7_10}

\bibitem{VanMark2021ModelMeetData}
Van Den~Brand, M., Cleophas, L., Gunasekaran, R., et~al.: Models meet data: Challenges to create virtual entities for digital twins (2021). \doi{10.1109/models-c53483.2021.00039}

\bibitem{Wen2024DecompDT}
Wen, X.J., Sun, Y.C., Liu, S.M., Bao, J.S., Zhang, D.: Fine-grained decomposition of complex digital twin systems driven by semantic-topological-dynamic associations. Journal of Manufacturing Systems  \textbf{77},  780--797 (2024)

\bibitem{Wu2022DTmultidispcollabDes}
Wu, Y.D., Zhou, L.Z., Zheng, P., Sun, Y.Q., Zhang, K.K.: A digital twin-based multidisciplinary collaborative design approach for complex engineering product development. Advanced Engineering Informatics  \textbf{52}, ~18 (2022)

\bibitem{yao2023systematic}
Yao, J.F., Yang, Y., Wang, X.C., Zhang, X.P.: Systematic review of digital twin technology and applications. Visual computing for industry, biomedicine, and art  \textbf{6}(1), ~10 (2023)

\bibitem{Zhou2024FedDTUAV}
Zhou, L., Leng, S., Wang, Q.: A federated digital twin framework for uavs-based mobile scenarios. IEEE Trans. Mobile Computing  \textbf{23}(6),  7377--7393 (2024). \doi{10.1109/tmc.2023.3335386}

\end{thebibliography}

\end{document}